\documentclass[a4paper,fleqn,usenatbib]{mnras}

\usepackage{newtxtext,newtxmath}

\usepackage[T1]{fontenc}
\usepackage{ae,aecompl}
\usepackage{hyperref}
\usepackage{cleveref}

\usepackage{graphicx}	
\usepackage{amsmath}	
\usepackage{amssymb}	

\title[Detection of the self-regulation of star formation]{Detection of the self-regulation of star formation in galaxy discs}

\author[J. Zaragoza-Cardiel et al.]{
Javier Zaragoza-Cardiel,$^{1,2}$\thanks{E-mail: javier.zaragoza@inaoep.mx}
Jacopo Fritz,$^{3}$ 
Itziar Aretxaga,$^{1}$ 
Divakara Mayya,$^{1}$ \newauthor
Daniel Rosa-Gonz\'alez,$^{1}$  
John E. Beckman,$^{4,5,6}$
Gustavo Bruzual, $^{3}$ 
Stephane Charlot  $^{7}$ \newauthor
and Luis Lomel\'i-N\'u\~nez  $^{1}$
\\
$^{1}$ Instituto Nacional de Astrof\'isica, \'Optica y Electr\'onica (INAOE), 
 Luis E. Erro 1, Tonantzintla, Puebla, C.P. 72840, M\'exico\\ 
$^{2}$ Consejo Nacional de Ciencia y Tecnolog\'ia, Av. Insurgentes Sur 1582, 03940, Ciudad de M\'exico, M\'exico\\
$^{3}$ Instituto de Radioastronom\'ia y Astrof\'isica, 
UNAM, Campus Morelia, 58089 Morelia, M\'exico\\
$^{4}$ Instituto de Astrof\'isica de Canarias, C/ V\'ia L\'actea s/n, 38205 La Laguna, Tenerife, Spain\\
$^{5}$ Department of Astrophysics, University of La Laguna, E-38200 La Laguna, Tenerife, Spain\\
$^{6}$ CSIC, 28006 Madrid, Spain\\
$^{7}$ Sorbonne Universit\'e, 
UPMC-CNRS, UMR7095, Institut d'Astrophysique de Paris, 
F-75014, Paris, France
}

\date{Accepted XXX. Received YYY; in original form ZZZ}

\pubyear{2019}

\begin{document}
\label{firstpage}
\pagerange{\pageref{firstpage}--\pageref{lastpage}}
\maketitle

\begin{abstract}

Stellar feedback has a notable influence on the formation and evolution of galaxies.
However, direct observational evidence is scarce. 
We have performed stellar population analysis using MUSE optical spectra of the spiral galaxy 
NGC 628 and 
find that current maximum star formation 
in spatially resolved regions is regulated 
according to the level of star formation in the recent past.
We propose a model based on 
the self-regulator or 
``bathtub'' models, but for spatially resolved regions of the galaxy. 
We name it the ``resolved self-regulator model'' and show that the predictions of this model are in 
agreement with the presented observations. 
We observe star 
formation self-regulation and estimate the mass-loading factor, 
$\eta=2.5 \pm 0.5$, consistent with values predicted by galaxy formation models. 
The method described here will help provide better constraints on those models.

\end{abstract}

\begin{keywords}
galaxies: evolution -- galaxies: formation -- galaxies: star formation -- galaxies: stellar content  
\end{keywords}



\section{Introduction}

Stellar feedback is commonly invoked to reproduce observed galaxy properties
in cosmologically based galaxy formation models 
\citep{2014MNRAS.445..581H,2014Natur.509..177V,2015MNRAS.446..521S}.
Theory supports stellar feedback as the main 
mechanism which regulates the star formation in low mass galaxies, while active galactic nuclear (AGN) feedback is predicted to
quench star formation in massive galaxies \citep{2012RAA....12..917S}. 
However, direct observational evidence supporting the stellar 
feedback scenario is lacking. 
Indirect evidence is based on 
the measurement of galaxy outflows 
\citep{2012MNRAS.426..801B,2014ApJ...792L..12K,2015ApJ...804...83S,2018MNRAS.tmp.2827L,2018A&A...619A.131B} or on 
analysing the chemical evolution of galaxies \citep{2017MNRAS.468.4494Z} rather than 
measuring directly the self regulation of star formation. 

The measurement of stellar feedback is critical in the reconciliation of the observed properties 
of galaxies with those from  
standard galaxy formation models \citep{2014MNRAS.445..581H,2014Natur.509..177V,2015MNRAS.446..521S}, 
and it is of great 
relevance in the stellar halo accretion rate efficiency \citep{2016MNRAS.455.2592R}, 
particularly in low mass galaxies. Unfortunately, estimates 
of the stellar feedback, are scarce and depend on the observability and 
uncertain mass of 
galactic winds \citep{2012MNRAS.426..801B,2014ApJ...792L..12K,2015ApJ...804...83S}.

Very recently, using star formation histories  
SFHs derived from stellar population analysis the 
AGN regulated star formation has been inferred, 
through the connection between star formation and black hole mass in massive galaxies \citep{2018Natur.553..307M}. 
The same technique can be used to find a connection between the star formation  at a given time, 
and star formation in the recent past. 
Although black hole activity is concentrated in the centre, it is capable 
of quenching star formation on galactic scales \citep{2005Natur.433..604D}. 
Star formation activity, however, although extended across the entire galaxy, produces its effects only close to its occurrence in space and time.

\begin{figure}
\begin{center}
\includegraphics[width=0.45\textwidth]{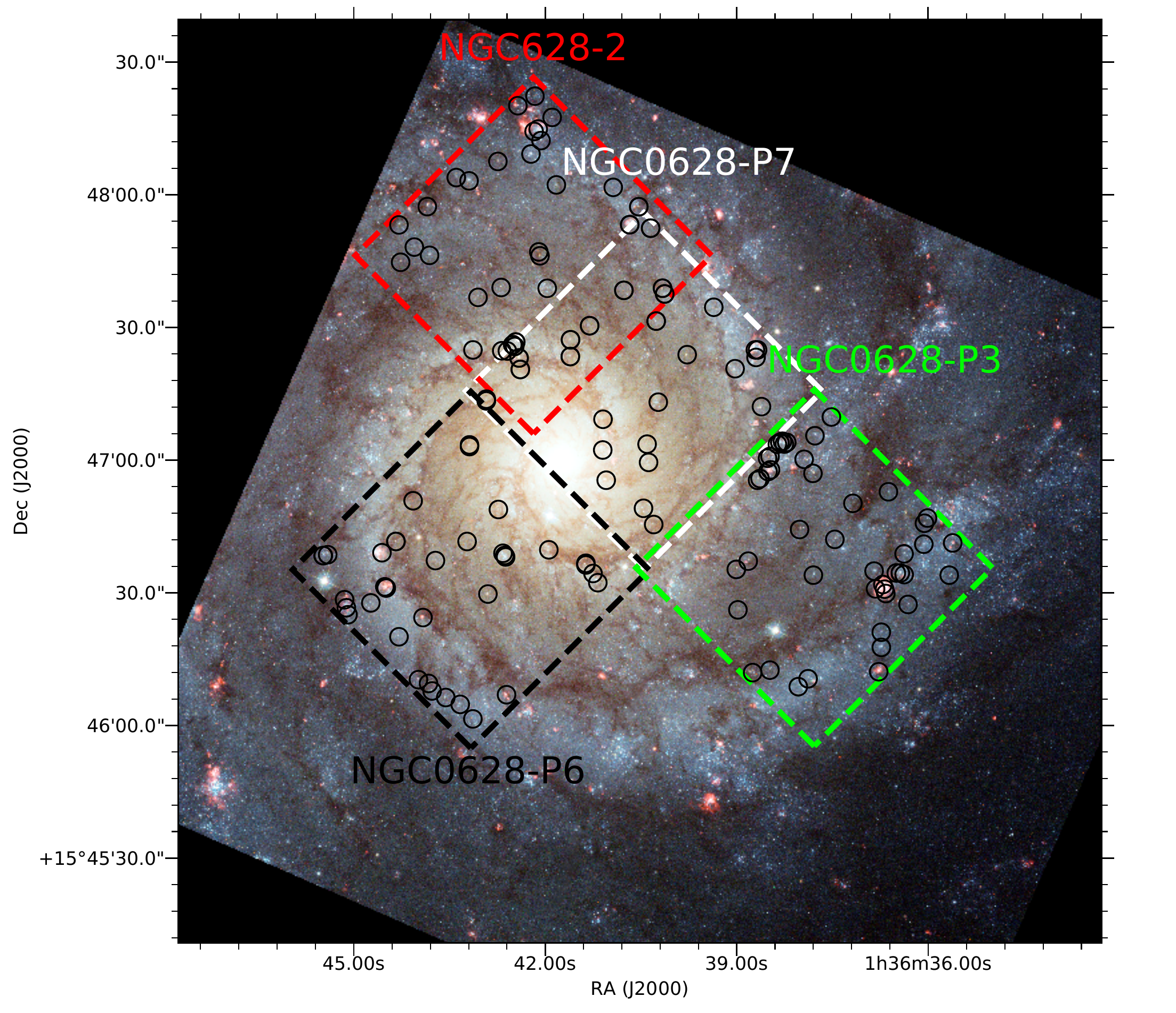} 
\end{center}
\caption{ 
Colour composite image from HST with $1\rm{arcmin}\times1\rm{arcmin}$ MUSE FOV  
overplotted.  There are 4 pointings. We label them with their ESO archive 
labels.  We also plot the compact star clusters centered regions as black circles.}
\label{fig_ngc628_muse}
\end{figure}

If we want to measure the effects of stellar feedback, we must focus on 
 time and spatial scales smaller than those affected by 
black hole feedback in massive galaxies, which is why
we focus here on looking for the past stellar feedback in spatially resolved regions in 
the galaxy NGC 628 using optical spectra. NGC 628 is an 
almost face-on, grand design spiral galaxy, with a     
$\rm{SFR}=0.15\thinspace\rm{M_{\odot}/yr}$ and a total stellar mass 
of $M_{*}=1.2\times10^{10}\thinspace\rm{M_{\odot}}$ 
\citep{2018ApJS..234...35Z}, right on the main sequence of star formation. 
NGC 628 has a disc-like pseudobulge 
\citep{2011AJ....142...16Z},  which helps us in the interpretation of the SFHs since the formation 
of the disc and the pseudobulge are both via secular evolution. The study would hardly 
be affected by the presence of a classical bulge since this would be made of stars older than the 
scale of ages we analyse.

We present the observational data and the stellar population synthesis analysis 
 in section \S2, while we propose a resolved self-regulation model of star formation 
  in section \S3. We compare the model with the observations in section \S4. 
Finally, we present our conclusions in section \S5.

\section{Observational data  \& Stellar population synthesis analysis}

We use data of NGC 628 from 
the Multi Unit Spectroscopic Explorer (MUSE) \citep{2010SPIE.7735E..08B}
mounted on the Very Large Telescope (VLT). The data were downloaded  
from the ESO 
archive\footnote{\url{http://archive.eso.org/wdb/wdb/adp/phase3_spectral/form?collection_name=MUSE}}. 
MUSE in its wide field mode is an integral field spectrograph of $1\rm{arcmin}\times1\rm{arcmin}$ 
field of view (FOV), 
with 2.6 \AA{} spectral resolution, seeing limited angular resolution, 
and it has a spectral range between 4750 and 9350 \AA.  

We used the pointings of NGC 628 shown in Fig. \ref{fig_ngc628_muse}: 
NGC0628-P3 (green square), NGC0628-P6 (black square), NGC0628-P7 (white square), 
and NGC628-2 (red square), with exposure times 
between $2350\rm{s}$ and $2780 \rm{s}$. 
We have used the publicly available flux and wavelength calibrated datacubes 
\citep{2016ApJ...827..103K,2017ApJ...834..174K,2018ApJ...863L..21K}.
NGC628-2 and NGC0628-P7 pointings overlap significantly, 
so we use the data only from 
the NGC628-2 pointing where there is no overlap with NGC0628-P7. 

In order to analyse the star formation regulation on different spatial scales, 
we divide each pointing into 9, 36, and 144 square regions, which results in 830, 415, and 207 pc-wide squares, respectively, as well as use circular regions of 
radius 87pc (black circles in Fig. \ref{fig_ngc628_muse}), centred on compact star clusters identified as described in \citet{2010MNRAS.405.1293S} using HST data (Lomel\'i-N\'u\~nez et al. in preparation).

After we extract the spectrum for each individual region, we 
correct it for Galactic extinction and mask the foreground stars.  
We fit the stellar continuum and the 
emission lines of these spectra using the 
SINOPSIS code\footnote{\url{http://www.crya.unam.mx/gente/j.fritz/JFhp/SINOPSIS.html}} 
\citep{2007A&A...470..137F,2017ApJ...848..132F}. 
SINOPSIS estimates the combination of single stellar population (SSP) models 
which best fits the equivalent widths of the  absorption and emission lines, and 
the stellar continuum in defined bands. 

 We do not assume any prior for the form of the SFH, allowing for a free form SFH.
We use the Calzetti dust attenuation law \citep{2000ApJ...533..682C}, 
and the Chabrier 2003 IMF \citep{2003PASP..115..763C} with stellar mass limits 
between $0.15\rm{M_{\odot}}$ 
and $120\rm{M_{\odot}}$.

We use the updated version of the Bruzual \& Charlot models described in \citet{2019MNRAS.483.2382W}. We use SSP
models for 3 metallicities ($Z=0.004$, $Z=0.02$, and  $Z=0.04$) 
in 12 age bins (2 Myr, 4 Myr, 7 Myr, 20 Myr, 57 Myr, 200 Myr, 
570 Myr, 1 Gyr, 3 Gyr, 5.75 Gyr, 10 Gyr, and 14 Gyr).
The emission lines are computed for the SSPs younger than 20 Myr \citep{2017ApJ...848..132F}
using the photoionisation code 
{\sc{Cloudy}} \citep{1993hbic.book.....F,1998PASP..110..761F,2013ApJ...767..123F}, 
assuming case B recombination \citep{1989agna.book.....O}, 
an electron temperature of $10^4 \rm{K}$, an electron density of $100\rm{cm}^{-3}$, 
and a gas cloud with an inner radius of $10^{-2} \rm{pc}$.

SINOPSIS fits the H$\alpha$ and H$\beta$ equivalent width estimates and 
the continuum in specific given bands. 
The continuum bands are chosen to  be unaffected by sky subtraction effects, specially  
in the red part of the spectrum.

The degeneracies 
between age, metallicity, and dust attenuation, are used by the SINOPSIS code to  
estimate the uncertainties in the derived parameters \citep{2007A&A...470..137F}.

We bin the SFH in four age bins, 
$20\rm{Myr}$, $570\rm{Myr}$, $5.7\rm{Gyr}$, 
and $14\rm{Gyr}$, and analyse the two most recent star formation events.  
Removing the two oldest bins in our analysis improves 
the confidence in our results, since they are worse constrained 
than the more recent bins. We also choose to be conservative, 
characterising the SFH in the last 570 Myr in the simplest way, 
by using only two age bins.  The SINOPSIS code has been 
tested with simulated and observed spectra \citep{2007A&A...470..137F,2011A&A...526A..45F}. 
SINOPSIS and 
similar codes have proved the validity of the use of SSP synthesis to 
extract non parametric SFHs in, at least, four age bins 
\citep{2005MNRAS.358..363C,2007A&A...470..137F,2011A&A...526A..45F,2016RMxAA..52...21S}.

We have restricted the analysis 
to a main sequence star forming disc galaxy, which has evolved through secular evolution 
in the studied time range (last 570 Myr).

\begin{figure*}

\begin{center}
\begin{tabular}{cc}

\includegraphics[width=0.35\textwidth]{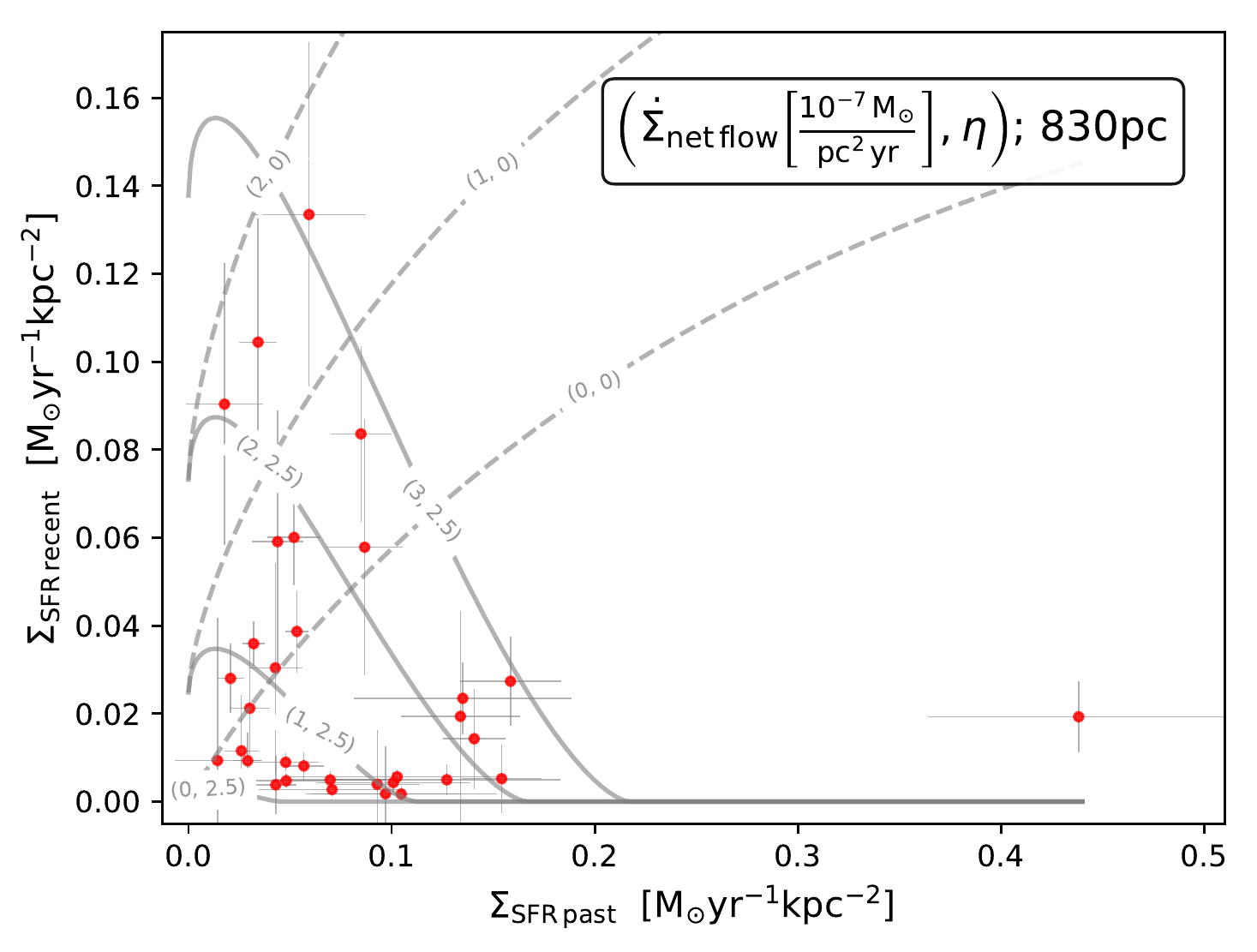}  
\put(-65,95){\large (a) }
&

\includegraphics[width=0.35\textwidth]{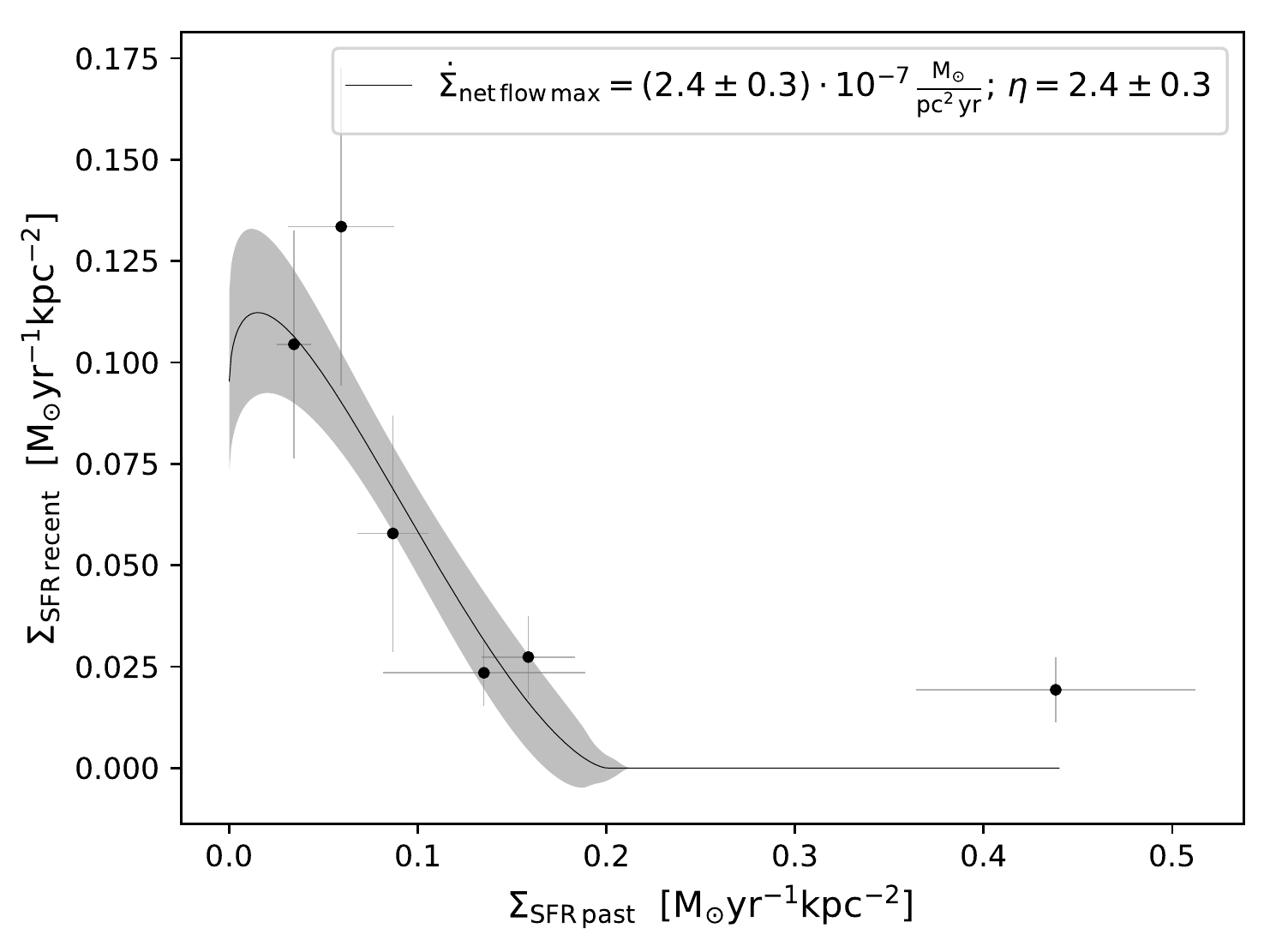}
\put(-65,95){\large (b) 830 pc}
\\

\includegraphics[width=0.35\textwidth]{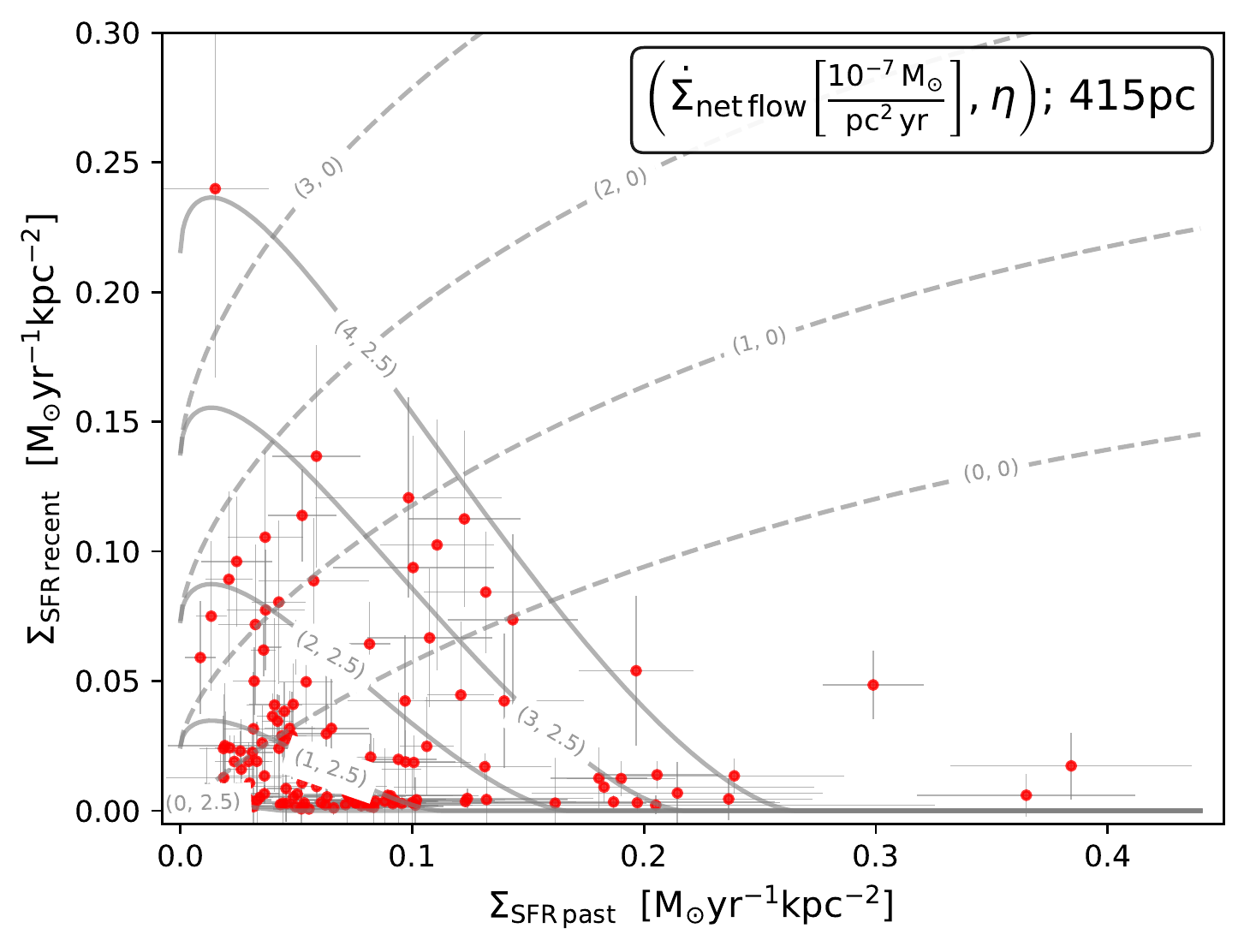}  
\put(-60,100){\large (c)}
&

\includegraphics[width=0.35\textwidth]{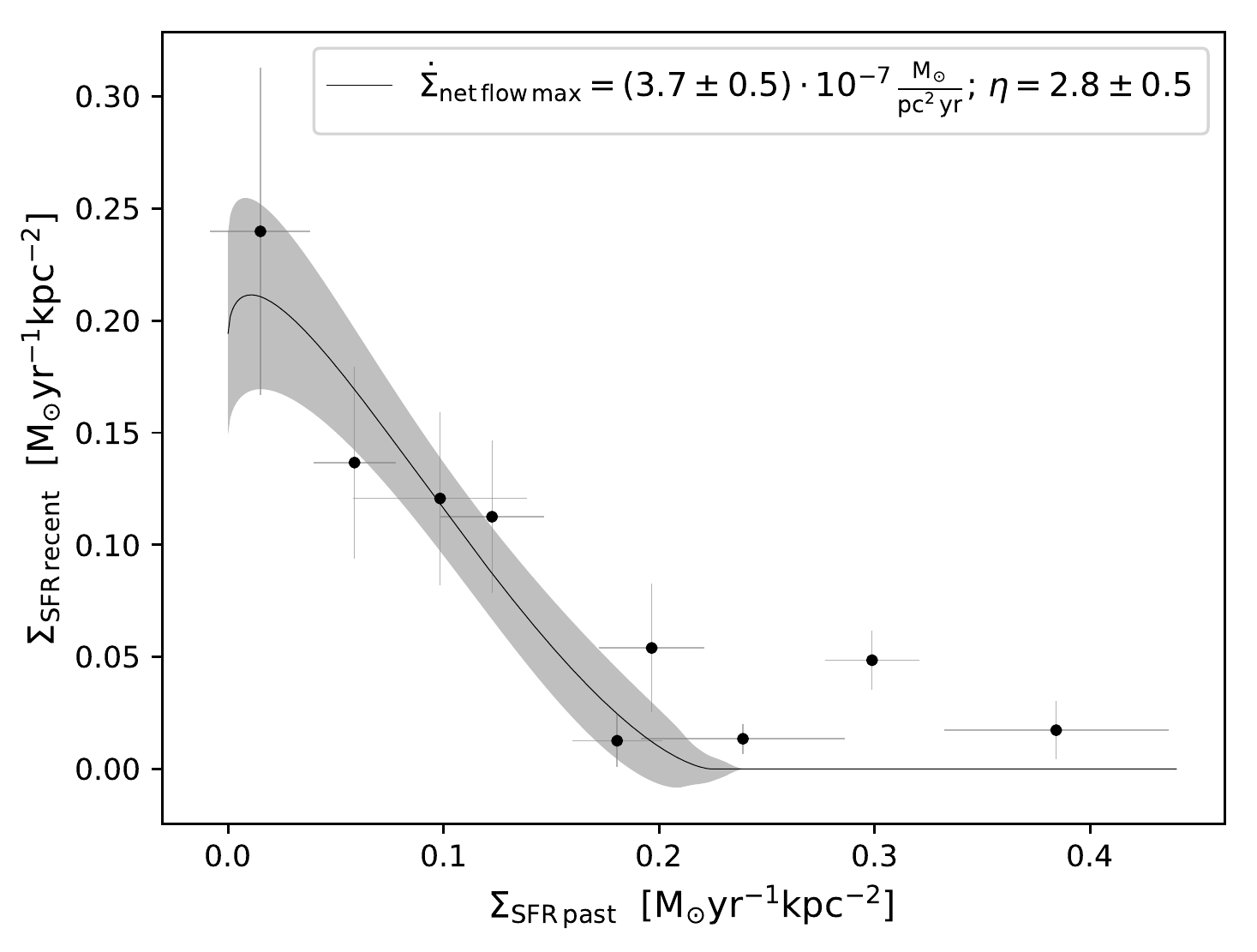}
\put(-60,100){\large (d) 415 pc}
\\

\includegraphics[width=0.35\textwidth]{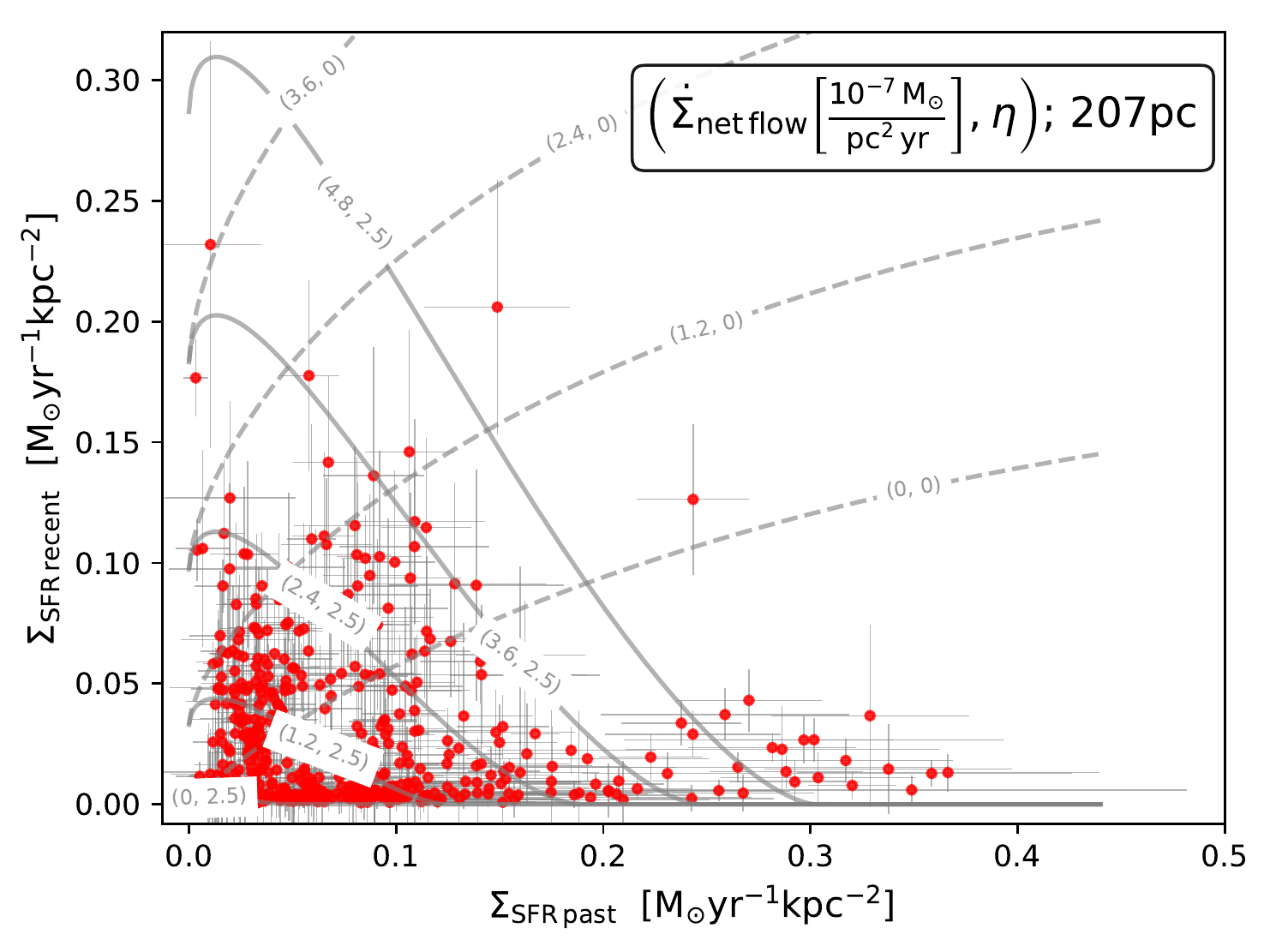}  
\put(-60,100){\large (e)}
&

\includegraphics[width=0.35\textwidth]{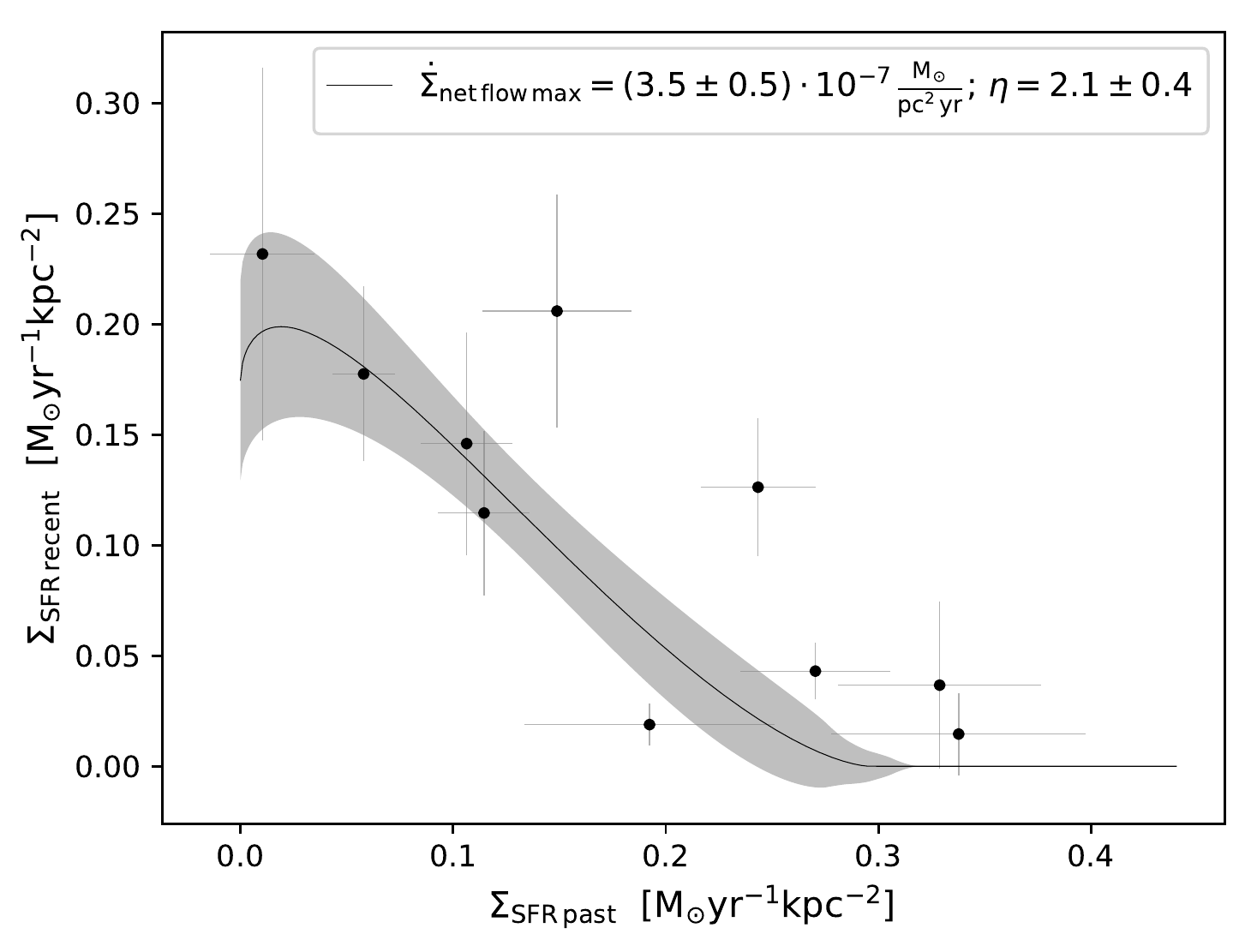}
\put(-60,100){\large (f) 207 pc}
\\

\includegraphics[width=0.35\textwidth]{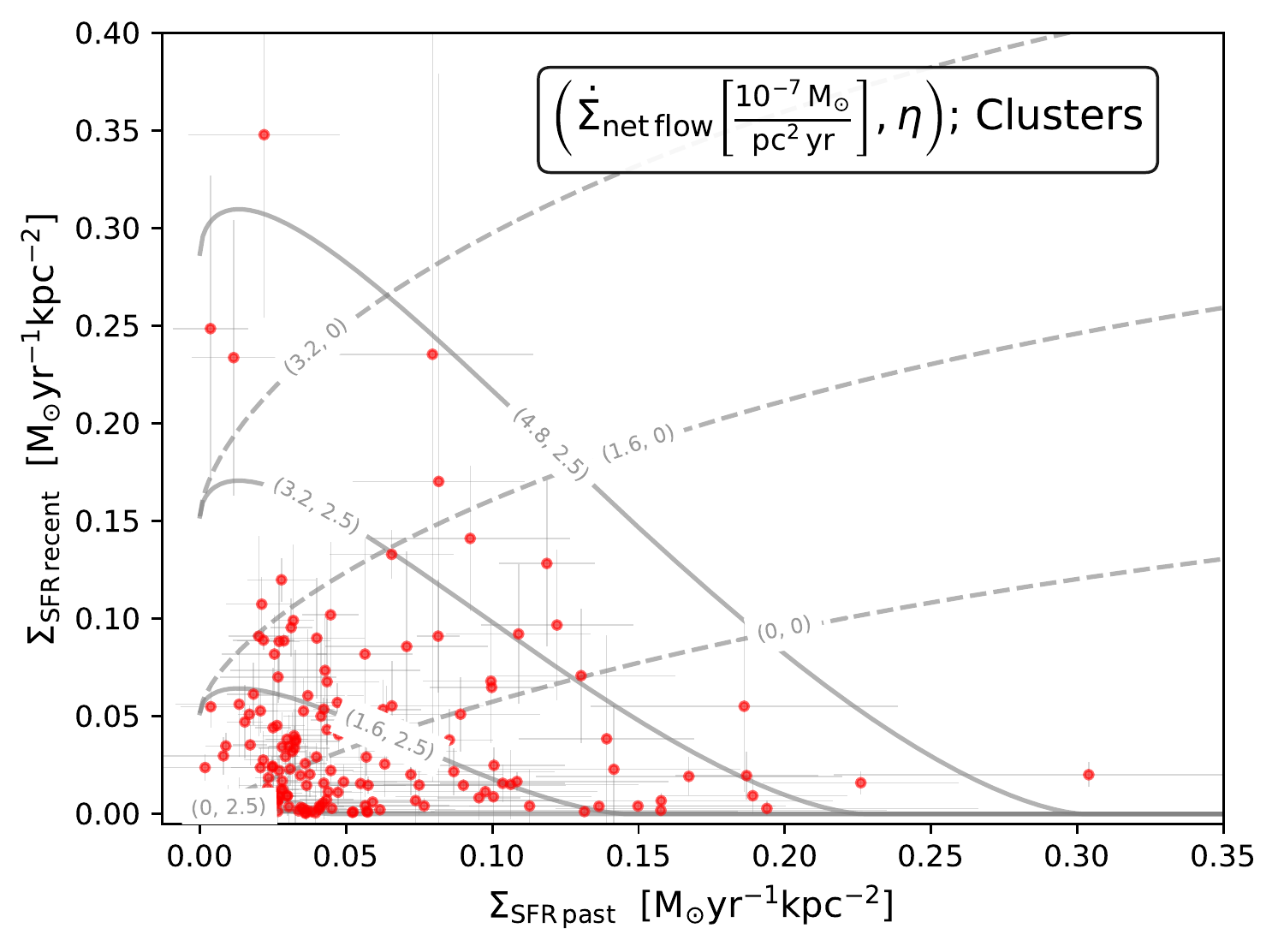}  
\put(-60,95){\large (g)}
&

\includegraphics[width=0.35\textwidth]{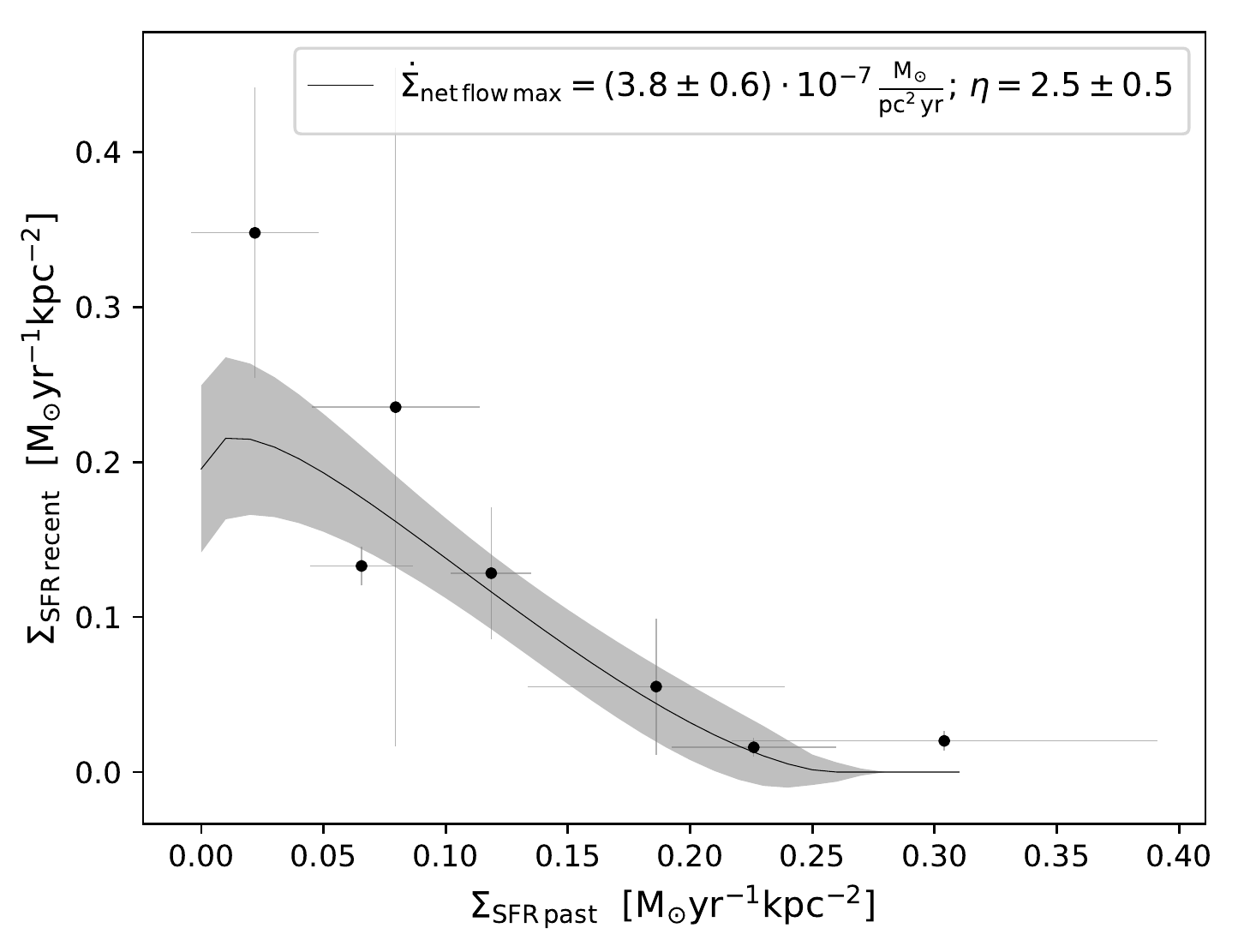}
\put(-60,100){\large (h) Clusters}
\\

\end{tabular}

\end{center}

\caption{
(a) SFR surface density in the 20 Myr age bin, $\Sigma_{\rm{SFR\thinspace recent}}$, versus 
SFR surface density in the 570 Myr age bin, $\Sigma_{\rm{SFR\thinspace past}}$ (red dots). 
 The proposed resolved self regulator model values (Eq. \ref{eq_etafit}) are plotted for several values of 
$\dot{\Sigma}_{\rm{net \thinspace flow}}$, 
and $\eta$ as solid ($\eta=2.5$) and dashed ($\eta=0$) grey lines. 
Each individual dot is for 830 pc-wide regions.
(b) SFR surface density in the 20 Myr age bin, $\Sigma_{\rm{SFR\thinspace recent}}$, versus 
SFR surface density in the 570 Myr age bin, $\Sigma_{\rm{SFR\thinspace past}}$, 
for those regions closest to maximum $\dot{\Sigma}_{\rm{net \thinspace flow}}$
(upper envelope).
The solid line is the fit of Eq. \ref{eq_etafit} to the envelope, 
where also we plot the 1-$\sigma$ uncertainty range of the fit as shadow regions.
Each individual dot is for an 830 pc-wide region.
(c) Panel is the same as (a) but for 415 pc-wide regions. 
(d) Panel is the same as (b) but for 415 pc-wide regions.
(e) Panel is the same as (a) but for 415 pc-wide regions. 
(f) Panel is the same as (b) but for 415 pc-wide regions. 
(g) Panel is the same as (a) but for circles of radius 87 pc and centred on star clusters.
(h) Panel is the same as (b) but for circles of radius 87 pc and centred on star clusters. 
}
\label{fig_sfh}
\end{figure*}

We extract the SFHs of each of the 207, 415, 
830 pc-wide regions, as well as those centered in the star clusers in four age bins. 
We analyse the current ($20\rm{Myr}$), and the recent past ($570\rm{Myr}$) SFHs bins, 
in order to check if they correlate, {\it{i. e.}}, if the 
current SFR is regulated by that in the recent past. 
For simplicity, we will refer to the recent past SFR (570 Myr age bin) as past SFR.

We compare the past SFR surface density, 
$\Sigma_{\rm{SFR\thinspace past}}$, 
with the recent SFR surface density (20 Myr age bin), $\Sigma_{\rm{SFR\thinspace recent}}$ 
in Fig. \ref{fig_sfh} (a, c, e,  and (g)) for the different region sizes as red dots.   

 Although we do not, strictly, observe a correlation between past and present SFR surface 
densities, we do find in Fig. 
\ref{fig_sfh} (a, c, e, and (g)) that for all the analysed scales, the maximum 
$\Sigma_{\rm{SFR\thinspace recent}}$ compared to the $\Sigma_{\rm{SFR\thinspace past}}$ is determined by the level of 
$\Sigma_{\rm{SFR\thinspace past}}$. Regions where $\Sigma_{\rm{SFR\thinspace past}}$ 
was the highest do 
not present $\Sigma_{\rm{SFR\thinspace recent}}$ as high as those with less $\Sigma_{\rm{SFR\thinspace past}}$, and viceversa, regions where $\Sigma_{\rm{SFR\thinspace recent}}$ 
is the highest, did not have $\Sigma_{\rm{SFR\thinspace past}}$ as high as those with less $\Sigma_{\rm{SFR\thinspace recent}}$. This can be explained, at least qualitatively,  if the maximum current SFR in a region is regulated by the SFR in the recent past.

\section{Resolved self-regulation star formation model}

\begin{figure}
\begin{center}
\includegraphics[width=0.45\textwidth]{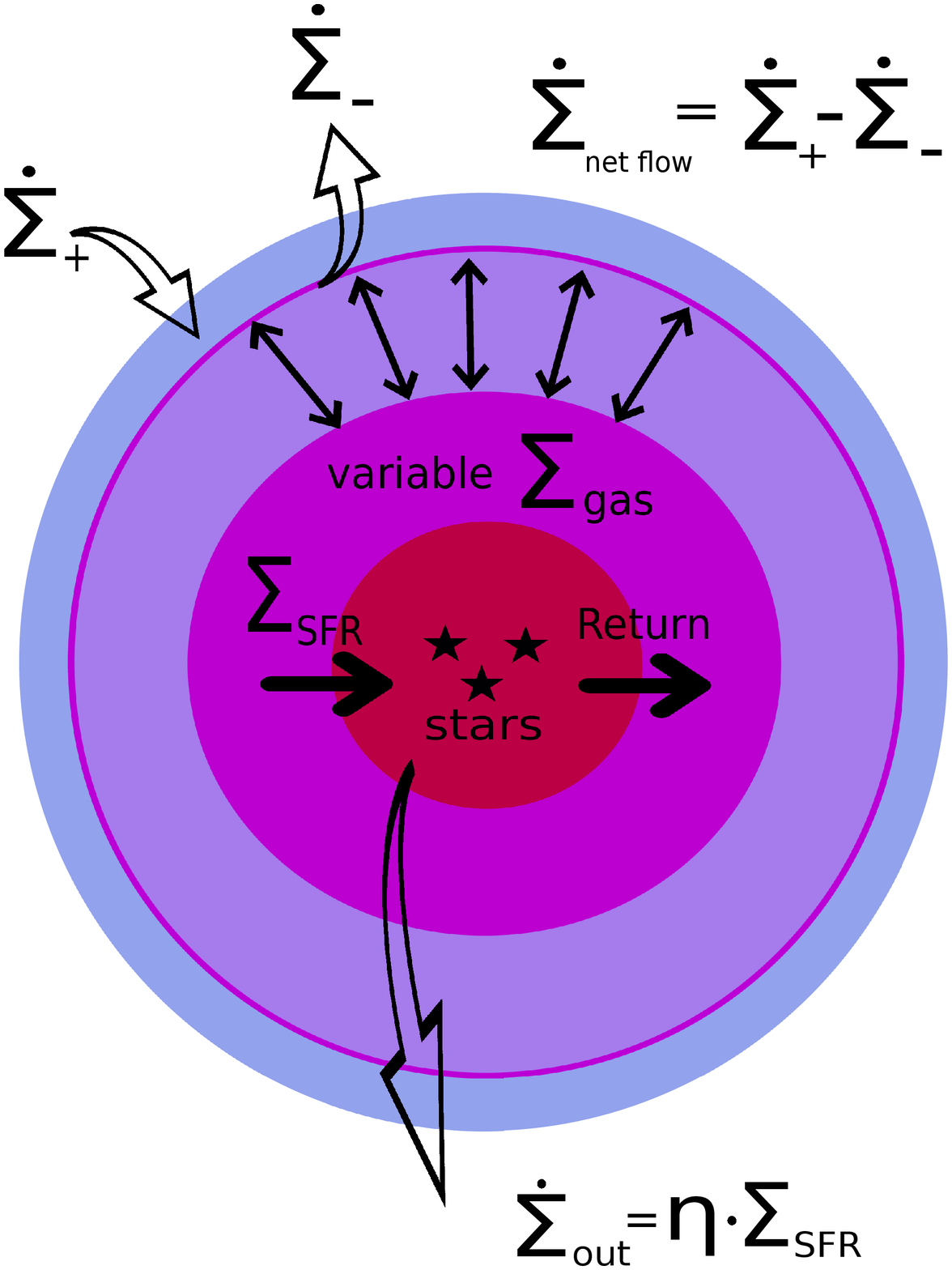} 
\end{center}
\caption{Resolved self-regulator model for a star forming region.  
The different layers within the circles refer to different components. 
The stellar disc and the star forming region  are co-rotating.
Invoking mass conservation, this model relates $\dot{\Sigma}_{\rm{gas}} $, $\dot{\Sigma}_{\rm{net \thinspace flow}}$, 
$\dot{\Sigma}_{\rm{out}}$, and $\Sigma_{\rm{SFR}}$.
}
\label{fig_regmod}
\end{figure}

In order to explain the observed star formation regulation, we propose 
a model based on the simplest models of star formation within galaxy evolution: 
the self-regulator or ``bathtub'' models 
\citep{2010ApJ...718.1001B,2013ApJ...772..119L,2014MNRAS.443.3643P,2014MNRAS.444.2071D,doi:10.1093/mnras/stu1142,2015MNRAS.448.2126A}. 
Briefly, these models assume mass conservation and relate, for a whole galaxy ,  
the change in time of the gas mass, 
$\dot{M}_{\rm{gas}}$, with 
the star formation rate, $\rm{SFR}$, 
the gas outflow rate, $\dot{M}_{\rm{out}}$, and the 
 gas inflow rate, $\dot{M}_{\rm{in}}$:

\begin{equation}
  \dot{M}_{\rm{gas}}={\dot{M}}_{\rm{in}}-{\rm{SFR}}(1-R+\eta)
 \label{eq_bath}
\end{equation}

\noindent where $R$ is the fraction of the mass that is returned to the interstellar medium, 
and $\eta$ is the ``mass loading 
factor'', 

\begin{equation}
 {\dot{M}}_{\rm{out}}=\eta{\rm{SFR}}.
 \label{eq_loading}
\end{equation}

 A resolved self-regulator model of the ISM has been already proposed \citep{2017MmSAI..88..533B}.
Based on the idea of a resolved version of the self-regulator models, we 
propose here one which is shown in the Fig. \ref{fig_regmod}.   
The model relates, assuming mass conservation, the change in time 
of the gas mass surface density, $\dot{\Sigma}_{\rm{gas}} $, to the SFR surface density, 
${\Sigma_{\rm{SFR}}}$, the net gas flow rate surface density, 
$\dot{\Sigma}_{\rm{net \thinspace flow}}$, and the gas outflow rate (due to stellar feedback) surface density, 
$\dot{\Sigma}_{\rm{out}}$:

\begin{equation}
  \dot{\Sigma}_{\rm{gas}}={\dot{\Sigma}}_{\rm{net \thinspace flow}}-{\Sigma_{\rm{SFR}}}(1-R+\eta)
   \label{eq_bath_res}
 \end{equation}

\noindent where 

\begin{equation}
 {\dot{\Sigma}}_{\rm{out}}=\eta{\Sigma_{\rm{SFR}}}.
 \label{eq_loading_res}
\end{equation}

The resolved self-regulator proposed here is for 
 an individual star or a group of stars which moves together through the rotation of the galaxy disc such as a massive star cluster ($> 500\rm{M_{\odot}}$ \citet{2003ARAA..41...57L}) . Thus, Eqs. \ref{eq_bath_res} and \ref{eq_loading_res} are defined locally, and 
the mass-loading factor is representative of massive star cluster scales ($\sim$pc \citet{2003ARAA..41...57L}). 
 The gas behaves as a collisional fluid, so 
we cannot assume that the gas follows 
the same rotation as the stars. For this reason we allow the gas to flow into and out of 
the region. We assume that these general flows are independent of any star formation inside the region, and that they  
depend on factors external to the region itself.  
The $\dot{\Sigma}_{\rm{net \thinspace flow}}$ term is defined as the difference 
between the gas surface density rate that enters the region, $\dot{\Sigma}_{+}$, minus 
the gas surface density rate that
leaves the region, $\dot{\Sigma}_{-}$. Finally, the gas that flows out 
of the region due to stellar feedback is quantified by the ${\dot{\Sigma}}_{\rm{out}}$ term 
and represents the part of the gas that flows out which depends on the star formation within the region.

We assume that large regions,such as the ones observed in this study, are a sum of 
individual star forming regions that obey Eq. \ref{eq_bath_res} for a specific 
 time interval. This means that 
star forming regions that produce unbound star clusters are split 
into individual stars, and star forming regions that produce bound clusters 
where the feedback is thought to be more important and more connected with the region 
properties, can be considered as a whole.
We do not consider the effect 
of feedback between regions, but only inside each region. We apply Eq. \ref{eq_bath_res} to 
a sum of star forming regions that we observe as a larger one at the present time: 
\begin{equation}
   \dot{\overline{{\Sigma}}}_{\rm{gas}}=\dot{\overline{{\Sigma}}}_{\rm{net \thinspace flow}}-\overline{\Sigma}_{\rm{SFR}}(1-R+\eta)
   \label{eq_bath_res_large}
\end{equation}

\noindent where the average is done in the whole region we observe. 
We can derive or 
approximate the average quantities from the observations reported in the previous section. 
We have measured the 
$\overline{\Sigma}_{\rm{SFR\thinspace past}}$, as the sum of individual star forming regions that are affecting the present measured $\overline{\Sigma}_{\rm{SFR\thinspace recent}}$. 
We convert $\dot{\overline{{\Sigma}}}_{\rm{gas}}$  into 
$ \frac{\overline{\Sigma}_{\rm{SFR\thinspace recent}}-\overline{\Sigma}_{\rm{SFR\thinspace past}}}{\Delta t}$.
We rewrite Eq. \ref{eq_bath_res_large} assuming  that the $\Sigma_{\rm{SFR}}$ and 
the $\Sigma_{\rm{gas}}$ are related by the Kennicutt-Schmidt (KS) law \citep{1998ApJ...498..541K}, 
$\Sigma_{\rm{SFR}}=A\Sigma_{\rm{gas}}^N$:

\begin{equation} 
\begin{split}
& \overline{\Sigma}_{\rm{SFR\thinspace recent}}= \\
&  A\left\{ \left[ \dot{\overline{\Sigma}}_{\rm{net \thinspace flow}}-\overline{\Sigma}_{\rm{SFR\thinspace past}}\left( 1-R+\eta \right)   \right]\Delta t +  \left[\frac{\overline{\Sigma}_{\rm{SFR\thinspace past}}}{A} \right]^{\frac{1}{N}} \right\}^N.
\end{split}
\label{eq_etafit}
\end{equation}

Eq. \ref{eq_etafit}, hence, relates the recent $\overline{\Sigma}_{\rm{SFR\thinspace recent}}$ 
with the past $\overline{\Sigma}_{\rm{SFR\thinspace past}}$, 
in terms of the mass loading factor, $\eta$, net gas flow rate surface density, 
$\dot{\overline{\Sigma}}_{\rm{net \thinspace flow}}$,  
the returned mass fraction, $R$, and the constants 
from the KS law, $A$, and $N$. 
 We assume the instantaneous recycling approximation \citep{2014ARA&A..52..415M} 
for stars more massive than $3\rm{M_{\odot}}$ ($\tau_{\rm{MS}}\sim 0.6\rm{Gyr}$), a Chabrier IMF 
\citep{2003PASP..115..763C}, 
and obtain a value of $R=0.27$. 

Deviations from the instantaneous recycling approximation, such as considering gas recycle from 
stars more massive than $1\rm{M_{\odot}}$ ($\tau_{\rm{MS}}\sim 10\rm{Gyr}$) or 
stars more massive than $4\rm{M_{\odot}}$ ($\tau_{\rm{MS}}\sim 0.3\rm{Gyr}$) adds 
an error of $\pm0.1$ in the considered value of $R$, which is a second order 
correction to the reported $\eta$ values.
Recycling gas from stars less massive than $1\rm{M_{\odot}}$  is negligible since their 
life times on main sequence are similar to or larger than the Hubble time.

We also assume, $A=-4.32$, and $N=1.56$ 
\citep{2007ApJ...671..333K}. 
 The gas which regulates the SFR is
 molecular and atomic gas, and hence 
the choice of $A$ and $N$ \citep{2007ApJ...671..333K}. 
This resolved self regulator model 
relates the gas inflow into the galaxy, mainly atomic gas, with that which is forming stars, mainly molecular gas. 
The two chosen ages for the recent and past age bins we used in our analysis are 20 Myr, and 570 Myr, 
respectively, so $\Delta t =550\thinspace \rm{Myr}$.

\section{Comparison between the self-regulator model and the observations}

We overplot in Fig. \ref{fig_sfh} (a, c, e and g) the theoretical  
relation between the past and recent SFR surface densities  assuming the 
 resolved self regulator model (Eq. \ref{eq_etafit}) 
 for a range of values of $\dot{\overline{\Sigma}}_{\rm{net \thinspace flow}}$  
and $\eta$.  
It is clear that the effect of increasing ${\dot{\overline{\Sigma}}}_{\rm{net \thinspace flow}}$ is to increase the 
recent SFR surface density, while the effect of $\eta$, 
is to reduce the recent SFR surface density in proportion to
the SFR surface density in the past. 
 
The case of no self-regulation ($\eta=0$) is 
 not favoured 
 by the combination of the model and observations presented here,   
since the $\eta=0$ case 
plotted as dashed grey lines in Fig. \ref{fig_sfh} show that  
the most intense star forming regions in the past, would not have 
as much ${\dot{\overline{\Sigma}}}_{\rm{net \thinspace flow}}$ as the less intense star forming 
regions in the past. 
There are no realistic mechanisms in the resolved self-regulator model 
that could produce larger net gas rate inflows into regions where there was not much star formation.

  For each different spatial scale, we can group the regions by their relative positions within the 
 resolved self regulator model family as a function of $\dot{\overline{\Sigma}}_{\rm{net \thinspace flow}}$ 
 (solid lines in Fig. \ref{fig_sfh} a, c, e and g). We identify   
  a group of regions closer to the model curve where $\dot{\overline{\Sigma}}_{\rm{net \thinspace flow}}$  is 
 maximum. 
 We define empirically those regions close to the curve where  
 $\dot{\overline{\Sigma}}_{\rm{net \thinspace flow}}$ is maximum 
 as those with maximum 
 $\overline{\Sigma}_{\rm{SFR\thinspace recent}}$ per bin of $\overline{\Sigma}_{\rm{SFR\thinspace past}}$. 
 We call this group of regions the upper envelope and plot them 
 as black dots in Fig.  \ref{fig_sfh} (b, d, f, and h) for the different region sizes.

Assuming stochasticity on $\dot{\overline{\Sigma}}_{\rm{net \thinspace flow}}$ inside the regions,  
the data are compatible with the resolved self-regulator star formation model with 
random flows inside a range, 
$\dot{\overline{\Sigma}}_{\rm{net \thinspace flow}}\in[\dot{\overline{\Sigma}}_{\rm{net \thinspace flow\thinspace min}},\dot{\overline{\Sigma}}_{\rm{net \thinspace flow\thinspace max}}]$. 
Although the value of $\dot{\overline{\Sigma}}_{\rm{net \thinspace flow}}$ could be negative, 
we can not distinguish between negative and zero 
$\dot{\overline{\Sigma}}_{\rm{net \thinspace flow}}$ 
since the effect of negative and zero net flow is the same, 
to quench the recent star formation. 

However, since there should exist a maximum net gas flow, 
$\dot{\overline{\Sigma}}_{\rm{net \thinspace flow\thinspace max}}$, there is a limit on the 
observed  $\overline{\Sigma}_{\rm{SFR\thinspace recent}}$. Assuming that 
in a normal star forming main sequence spiral galaxy like NGC 628, there are 
a few regions where in 550Myr their net gas flow is approximately $\dot{\overline{\Sigma}}_{\rm{net \thinspace flow\thinspace max}}$, we can approximate those regions as those identified on the upper 
envelope.

\begin{table}
	\centering
	\caption{Resolved self-regulator model (Eq. \ref{eq_etafit}) fit results. 
	 The fits are done to the regions on the upper envelope (plotted as black dots in Figs. \ref{fig_sfh} b, d, f, and h), which we interpret as region 
	where $\dot{\overline{\Sigma}}_{\rm{net \thinspace flow\thinspace}}
	\approx\dot{\overline{\Sigma}}_{\rm{net \thinspace flow\thinspace max}}$}
	\label{table_fits}
	\begin{tabular}{lcc} 
		\hline

Region size (pc) & $\dot{\overline{\Sigma}}_{\rm{flow\thinspace max}}$ ($10^{-7}\thinspace\rm{\frac{M_{\odot}}{pc^2\thinspace yr}}$)  & $\eta$ \\

		\hline
		$830$ & $2.4\pm0.3$ & $2.4\pm0.3$ \\
		$415$ & $3.7\pm0.3$ & $2.8\pm0.5$ \\
		$207$ & $3.5\pm0.3$ & $2.1\pm0.4$ \\
		$87$ & $3.8\pm0.6$ & $2.5\pm0.5$ \\
		\hline
	\end{tabular}
\end{table}

We use  the resolved self regulator model 
(Eq. \ref{eq_etafit}) to fit the data (error weighted) from the envelopes and estimate 
the loading factor, $\eta$. We report the results in Tab. \ref{table_fits}

 Consistent values of $\eta$ between different spatial scales imply that $\eta$ is 
  unique,  
independent of the analysed area, and representaitve of a local individual star forming region.

\section{Conclusions}

 We presented an analysis of the evolution in time of the star formation 
in resolved regions and compact star clusters of the NGC 628 galaxy. 
On all the analyzed scales, we find that the 
maximum $\Sigma_{\rm{SFR\thinspace recent}}$  is regulated by the 
$\Sigma_{\rm{SFR\thinspace past}}$.
The proposed model based on the self-regulator star formation model,   
but for individual star forming regions, is compatible with 
the reported observations. 
We report the measurement of how the star formation regulates itself 
on a 550Myr time-scale. 
This measurement of the stellar feedback 
suggests that the star formation does regulate itself inside galaxies, as 
required by cosmologically based galaxy formation models.

 We find that 
$\eta$ is independent of the chosen studied scale within the 
reported uncertainties, which is in agreement with the definition 
of $\eta$ as representative of individual star forming regions.
The fact that we find the same $\eta$ for regions centered on compact 
star clusters, where the stellar populations should be more connected 
on the analyzed time scale since compact star clusters are more bound, 
probably means that the assumptions made 
 about the comparison between the the self-regulator model and the data 
 are broadly correct.
In that case the estimated mass-loading factor is also representative  
of the galaxy. 
The $\eta=2.5\pm0.5$ value of the smallest scales and estimated from 
regions centred on compact star clusters is probably the most 
representative of the galaxy, and it is consistent 
with what is found indirectly \citep{2012MNRAS.426..801B}, 
and with what $\Lambda$-CDM framed galaxy models predict  
\citep{2012MNRAS.421.3522H,2016MNRAS.455.2592R} for 
an $L_{*}$ galaxy such as NGC 628.

With the new method presented here we plan to estimate 
the mass loading factor for different galaxies and analyse 
the self regulation of star formation as a function of different galaxy properties 
to compare, for example,  
with the prediction that the smaller the mass of 
the galaxy, the larger is $\eta$.

\section*{Acknowledgements}
 The authors thank the anonymous referee whose comments have led to signifcant improvements in the paper.
 JZC and IA's work is funded by a CONACYT grant through project FDC-2018-1848. 
  GB acknowledges financial support through PAPIIT project IG100319 from DGAPA-UNAM.
 This research has made use of the services of the ESO Science Archive Facility, 
 Astropy,\footnote{\url{http://www.astropy.org}} a 
 community-developed core Python package for Astronomy \citep{2013A&A...558A..33A,2018AJ....156..123A}, 
 and APLpy, 
 an open-source plotting package for Python \citep{2012ascl.soft08017R}.
 Based on observations collected at the European Southern Observatory 
 under ESO programmes 094.C-0623(A), and 098.C-0484(A).

 %
%

%
\bibliographystyle{mnras}


\end{document}